\documentclass[fleqn,twoside]{article}
\usepackage{espcrc2}

\usepackage{graphicx}
\usepackage[figuresright]{rotating}

\newcommand{\AmS}{{\protect\the\textfont2
  A\kern-.1667em\lower.5ex\hbox{M}\kern-.125emS}}

\title{Gallium data variability and KamLAND}

\author{\underline{Jo\~{a}o Pulido}, \address[CFTP]{Centro de F\'{\i}sica Te\'{o}rica das Part\'{\i}culas,
        Avenida Rovisco Pais P-1049-001, Lisboa, Portugal\\
        }%
        \thanks{email: pulido@cftp.ist.utl.pt}
        Bhag Chauhan, \addressmark      Marco Picariello
        \address{Dipartimento di Fisica, 
        Universit\`{a} di Lecce, \\
        Via Arnesano, ex Collegio Fiorini, I-73100 Lecce, Italia
        }}
\begin{document}

\begin{abstract}
We present an alternative fit to the conventional solar+KamLAND one which 
takes into account the possible time dependence of the Ga data and relies 
on the partial conversion of active $\rightarrow$  sterile neutrinos via the 
magnetic moment/solar field interaction. We evaluate the prediction for the 
solar neutrino rates obtaining a fit of similar quality as the
LMA one. We also evaluate the KamLAND antineutrino survival probability as
a function of reactor distance and find a better agreement with data as
compared to LMA. 
\vspace{1pc}
\end{abstract}

\maketitle

%\section{FORMAT}

{\bf 1.} Time modulation of the solar neutrino flux is probably the most 
important issue after LMA has been asserted as the dominant solution to 
the solar neutrino problem. Many efforts have been undertaken recently 
by the experimental collaborations and theoretical groups
\cite{Aharmim:2005iu,Yoo:2003rc,Sturrock:2004jv} to look into modulation.
If this is confirmed it will probably imply the existence of a sizable 
neutrino magnetic moment $\mu_{\nu}$ and hence a wealth of new physics. 
The idea of neutrinos interacting with the solar magnetic field through
their magnetic moment was first introduced in 1970 \cite{Cisneros},
revived in 1986 in connection to solar activity \cite{Okun:1986na} and 
later viewed in terms of a resonant mechanism - the resonant spin flavour 
precession (RSFP) \cite{Lim:1987tk,Akhmedov:1988uk}.
In summary active $\nu_e's$ produced in the sun are assumed to be converted 
to sterile ones owing to the above interaction so that at times of intense 
solar activity a strong field leads to a large conversion with little or
no conversion otherwise. Hence a neutrino flux anticorrelated to solar 
activity. 

{\bf 2.} The Gallium solar neutrino data have been consistently decreasing 
in time, a striking fact that may be the signature of a long term periodicity. 
In fact, as can be seen from table 1, there is a 2.4$\sigma$ discrepancy in 
the combined results over the two periods which may be an indication of a 
possible anticorrelation of the Ga rate with the 11-year solar cycle. Ga 
are the only experiments with a significant contribution from $pp$, $^7 Be$
neutrinos (together they account for $\simeq$ 80\% of the event rate and
more than 99\% of the total solar flux). No other experiment shows such a
variational effect, so the time dependence of the low energy fluxes
becomes an open possibility.  On the contrary, averaging Ga rates completely 
over time may erase important information already contained in the data. 
Hence we propose an alternative to the conventional solar+KamLAND fit 
attempting instead at two separate fits to the two Ga data sets consistent 
with all other solar and KamLAND data.

\begin{table*}[htb]
\caption{Average rates for Ga experiments in SNU 
}
\label{table:1}
\newcommand{\m}{\hphantom{$-$}}
\newcommand{\cc}[1]{\multicolumn{1}{c}{#1}}
\begin{tabular}{@{}lll}
\hline
Period & 1991-97 & 1998-03    \\
\hline
%b 
SAGE+Ga/GNO   & $77.8\pm5.0$ & $63.3\pm3.6$ \\[2pt]
Ga/GNO only   & $77.5\pm7.7$ & $62.9\pm6.0$ \\[2pt]
SAGE only     & $79.2\pm8.6$ & $63.9\pm5.0$ \\[2pt]
av. no. of suspots & 52  & 100 \\ [2pt]
\hline
\end{tabular}\\[2pt]
\end{table*}

%\begin{figure}[htb]
%%\vspace{6pt}
%\vspace{9pt}
%\hspace*{-1.6cm}
%%\framebox[55mm]{\rule[-21mm]{0mm}{43mm}}
%\includegraphics*[angle=270,width=25pc]{p_osc_now.ps}
%%\includegraphics*[angle=270,width=22pc]{p_osc_now.ps}
%\caption{Survival probabilities for LMA (solid line) and our model parameters 
%[eqs.(4),(5)] (dashed line).}
%\label{fig:largenenough}
%\end{figure}

{\bf 3.} Our model \cite{Chauhan:2004sf,Chauhan:2005pn} introduces 
light sterile neutrinos which only communicate 
to active ones through one magnetic moment interaction, so that in 
the vacuum

\begin{equation}
\left(\begin{array}{c}\nu_{s}\\ \nu_{e}\\ \nu_{x}\end{array}\right)=
\left(\begin{array}{ccc}1&0&0\\ 0&
c_{\theta}&s_{\theta}\\ 0&-s_{\theta}&
c_{\theta}\end{array}\right)\left(\begin{array}{c}\nu_{0}\\ \nu_{1}\\
\nu_{2}\end{array}\right)
\end{equation}
and in matter 

\begin{equation}
\cal{H}_{\rm {M}}\!\!=\!\!\left(\begin{array}{ccc}\frac{-\Delta m^2_{10}}{2E}&
\mu_{\nu}B&0 \\ \mu_{\nu}B& \frac{\Delta m^2_{21}}{2E}s^2_{\theta}+V_e&
\frac{\Delta m^2_{21}}{4E}s_{2\theta}\\ 0&\frac{\Delta m^2_{21}}{4E}s_{2\theta}&
\frac{\Delta m^2_{21}}{2E}c^2_{\theta}+V_{x}\end{array}\right)
\end{equation}
where in (1) $\nu_{S}$ denotes the sterile neutrino and $V_e,~V_{x}$ are the
refraction indices. The parameter 
$\Delta m^2_{10}=m^2_{1}-m^2_{0}$ dictates the location of the active $\rightarrow$
sterile transition. It is plausible that the solar magnetic activity, as evidenced 
by sunspots, may extend down to the bottom of the convective zone at $x=0.71$ of
the solar radius. Furthermore
it is at the tachocline (the region extending from approximately $x=0.68$ to $x=0.72$) 
that the magnetic field is supposed to be peaked, this peak 
being closely connected with the local maximum of the angular velocity gradient.
Therefore a 'plausible' magnetic field will be peaked around $x=0.71$
as shown in fig.1
of \cite{Chauhan:2005pn}, and we assume the height of this peak to be closely
connected to the solar activity. Since maximum conversion occurs near the
resonance, low energy neutrinos are expected to resonate around this region and
their flux may reflect the solar magnetic activity. Hence we need $\Delta m^2_{10}=
O(10^{-8}eV^2)$, thus excluding conversion to active neutrinos for which both
possible values of the mass square difference are larger. The large order of 
magnitude discrepancy between $\Delta m^2_{10}$ and the other two implies
the LMA and the spin flavour precession resonance to be located far apart, thus
precluding their interference.

{\bf 4.} For typical parameter values 
\begin{equation}
\label{B_peak}
B_{peak}=280kG
\end{equation}
\begin{equation}
\label{dm_10}
\Delta m^2_{10}=-1.7\times 10^{-8}eV^2 
\end{equation}
\begin{equation}
\label{dm21}
\Delta m^2_{21}=8.2\times 10^{-5}eV^2 
\end{equation}
\begin{equation}
\label{Kam}
tan^2 \theta=0.31
\end{equation}
we obtain the predictions shown in table 2 where the first two rows refer to the
two consecutive periods in table 1. We made use of the comparatively large
uncertainty in the $^7 Be$ flux from solar models \cite{Serenelli} and 
assumed the normalizing factor $f_{Be}=1.1$. For comparison we recall the
LMA best fit parameters for KamLAND only
\begin{equation}
\Delta m^2_{21}\!\!=\!\!7.9\pm^{0.6}_{0.5}\!\times 10\!^{-5}\!eV\!^2,
tan^2 \theta\!\!=\!\!0.46\pm^{4.5}_{0.25} (2\sigma).
\end{equation}

\begin{table*}[htb]
\caption{The two sets of rates for Ga experiments in SNU. For
comparison we list the LMA case.}
\label{table:2}
\newcommand{\m}{\hphantom{$-$}}
\newcommand{\cc}[1]{\multicolumn{1}{c}{#1}}
\begin{tabular}{@{}llllllllll}
%\hline
%Period & 1991-97 & 1998-03    \\
\hline
& 
Ga & Cl & K (SK) & $\!\!\rm{SNO_{NC}}\!\!$ &$\!\!\rm{SNO_{CC}}\!\!$&$\!\!\rm{SNO_{ES}}\!\!$
& $\!\!\chi^2_{rates}\!\!$ & $\chi^2_{{SK}_{sp}}$ & $\chi^2_{{SNO}_{gl}}$\\ \hline
%& $\chi^2_{KL}$\\ \hline
Period 1991-97 &
73.8 &  2.72 &  2.29 &  &  &  & 2.78 &  &   \\
Period 1998-03  &
60.3 &   &  2.28 & 5.65 & 1.59 & 2.25 & 0.54 & 47.5 & 48.5 \\ 
LMA &
64.8 &  2.74 &  2.30  &  5.10 & 1.75 & 2.28  & 0.95 & 45.7 & 43.1 \\ \hline
%\hline
\end{tabular}\\[2pt]
\end{table*}

{\bf 5.} Comparison of table 2 and eqs.(3)-(6) with fig.4 
of ref.\cite{Araki:2004mb} shows that the fit presented here lies well
within the 95\% CL of the KamLAND best fit but just outside the 99.73\%CL
of the solar fit. Such a result is to be expected since we neglected the 
conventional solar fit. We have for the global $\chi^2$ (excluding KamLAND):
\begin{equation}
\chi^2=96.6/94~d.o.f.~,~\chi^2_{_{LMA}}=89.8/93~d.o.f.
\end{equation}
We have also evaluated the antineutrino survival probability for our model 
parameters $\Delta m^2_{21}$ and $tan^2 \theta$ as given above [eqs.(4),(5)]as a 
function of effective reactor distance and compared it with the standard LMA one. 
This is given by the well known expression
\begin{equation}
P_{osc}(E_{\bar\nu},L)=1-sin^2 \theta sin^2 \left(\frac{\Delta m^2_{21}L}{4E_{\bar\nu}}
\right)
\end{equation}
and the results are shown in fig.1. For the average reactor distance to which the
data refer (180 km) \cite{Araki:2004mb} we find, as seen from fig.1, $P=0.628$ and
for the LMA case $P_{LMA}=0.578$ while the KamLAND collaboration quotes 
$P=0.658\pm0.044(stat)\pm0.047(syst)$.   

{\bf 6.} To conclude, we have investigated the consequences of a possible 
decrease of the Gallium event rate consistently observed by the two Gallium
experiments as a function of time and related this
decrease to solar activity. Our main motivation arises from the fact that
no other experiment exhibits such a data variation and the Gallium event rate
is the only one with a strong dependence on the low energy (LE) neutrino flux from the
sun. Investigating the variability of the solar neutrino flux is the most important 
challenge facing us in solar neutrino physics, now that LMA is known to play a 
major role in the solution to the solar neutrino problem. A clear distinction between 
our scenario and the LMA one will 
only be possible either with data improvement from KamLAND or average reactor
distances below 110-120km as can be seen from fig.1, or both. Interestingly enough
our scenario based on LMA+RSFP gives an antineutrino survival 
probability at the average reactor distance of 180 km in better agreement with data 
than the LMA case. Important evolutions in KamLAND are expected: new reactors may come 
into operation while others cease and fluxes almost constantly change. This will
provide us new opportunities by changing effective reactor distance thereby 
providing a distinction between the two scenarios investigated. Also the variability 
of the LE solar neutrino flux, possibly in connection to solar activity, will also be 
tested by the forthcoming LE experiments: KamLAND (solar mode), LENS, Borexino and SNO+.

\begin{figure}[htb]
%\vspace{6pt}
\vspace{9pt}
\hspace*{-1.0cm}
%\framebox[55mm]{\rule[-21mm]{0mm}{43mm}}
\includegraphics*[angle=270,width=25pc]{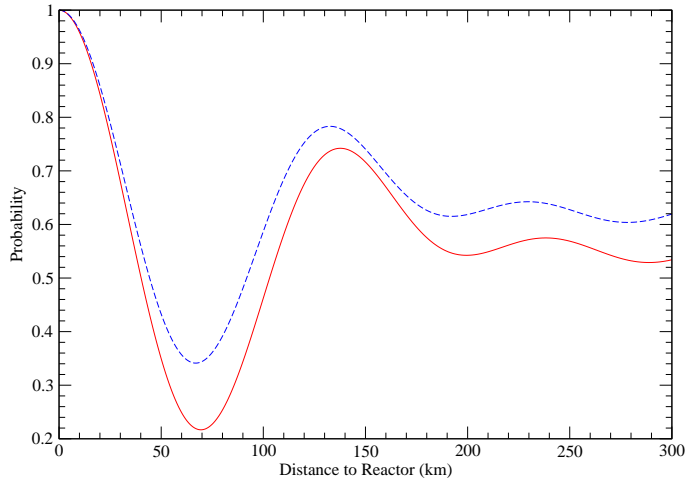}
\caption{Survival probabilities for LMA (solid line) and our model parameters 
[eqs.(4),(5)] (dashed line).}
\label{fig:largenenough}
\end{figure}

\end{document}